\documentclass[twocolumn]{aa}
\usepackage{psfig}
\begin{document}


\title{Distance to the Active Galaxy NGC~6951 via the Type~Ia
Supernova 2000E\footnote[1]{Based on observations obtained at German-Spanish
Astronomical Centre, Calar Alto, operated by the Max-Planck-Institute 
for Astronomy, Heidelberg, jointly with the Spanish National 
Commission for Astronomy}}

\author{J.Vink\'o \inst{1} \and 
B.Cs\'ak \inst{1,6} \and
Sz.Csizmadia \inst{2} \and
G.F\H ur\'esz \inst{3,7} \and 
L.L.Kiss \inst{3,7} \and
K.S\'arneczky \inst{4,6,7} \and
Gy.Szab\'o \inst{3,6,7} \and
K.Szil\'adi \inst{3,7} \and
I.B.B\'{\i}r\'o \inst{5,7}}

\institute{Department of Optics \& Quantum Electronics, University of Szeged,
POB 406, Szeged, H-6701 Hungary \and
Konkoly Observatory of the Hungarian Academy of Sciences, POB 67, 
Budapest, H-1525 Hungary \and
Department of Experimental Physics, University of Szeged, D\'om t\'er 9., Szeged,
H-6720 Hungary \and
Department of Physical Geography, ELTE University, 
Ludovika t\'er 2., Budapest, H-1088 Hungary \and
Baja Observatory, POB 766, Baja, H-6500 Hungary \and 
Visiting Astronomer, German-Spanish Astronomical Centre,
Calar Alto, Spain \and
Guest Observer, Piszk\'estet\H o Station, 
Konkoly Observatory, Hungary}

\titlerunning{Distance to NGC~6951 via SN~2000E}
\authorrunning{J.Vink\'o {\it et al.}}
\offprints{J. Vink\'o, \\ 
\email{vinko@physx.u-szeged.hu}}
\date{}


\abstract{
CCD-photometry and low-resolution spectroscopy 
of the bright supernova SN~2000E in NGC~6951 are
presented. Both the light curve extending up to
150 days past maximum and the spectra obtained
at 1 month past maximum confirm that SN~2000E
is of Type Ia.
The reddening of SN~2000E is determined as 
$E(B-V) = 0.36 \pm 0.15$, its error is
mainly due to uncertainties in the predicted
SN $(B-V)$ colour at late epochs.
The $V (RI)_{\rm C}$ light curves are analyzed
with the Multi-Colour Light Curve Shape (MLCS)
method. The shape of the late light curve 
suggests that SN~2000E was overluminous by about
0.5 mag at maximum comparing with a fiducial
SN Ia. This results in an updated distance 
of 33 $\pm$ 8 Mpc of NGC~6951 (corrected
for interstellar absorption). The SN-based
distance modulus is larger by about +0.7 mag
than the previous Tully-Fisher estimates.
However, possible systematic errors due to
ambiguities in the reddening determination
and estimates of the maximum luminosity
of SN~2000E may plague the present distance 
measurement.
}

\maketitle

\section{Introduction}

Type Ia supernovae (SNe Ia) are considered to be the most
reliable distance indicators on extragalactic, even cosmological 
distance scales (e.g. \cite{key1,par,perl1,riess2,hamuy1}
and references therein). This is mainly based on their exceptional
brightness and homogeneity, despite of the existence of ``peculiar''
SNe~Ia, such as SN 1991T or SN 1991bg (e.g. \cite{filip1}). Although
the frequency of these ``standard bombs'' (\cite{jha}) is low,
regular monitoring of numerous galaxies (e.g. by the Lick Observatory 
Supernova Search, the Nearby Galaxies Supernova Search, etc.) 
supplies more than a hundred Type Ia SN events per year. 
The reliability of SN-based distances
is increased by the number of bright, well-observed, nearby
SNe~Ia in host galaxies whose distances can also be determined by
other methods, such as Cepheids (\cite{saha,key1,jha}), Tully-Fisher
relation, or surface brightness fluctuation (e.g. \cite{riess1}).

In this paper we present an updated distance to the type 2 Seyfert
galaxy NGC~6951 via the Type Ia SN~2000E. This galaxy has received
considerable attention recently, especially its active nucleus
and circumnuclear star-forming ring (\cite{boer,barth,elm,kohno,perez}).
Its distance has been determined via Tully-Fisher relation by
several groups. \cite{bott} gives $\mu_{\rm 0} = 31.85$ mag for the
true distance modulus (corresponding to 23.4 Mpc), while 
\cite{tully} lists 24.1 Mpc ($\mu_{\rm 0} = 31.91$ mag). 

SN~2000E has occurred just outside the origin of the long, northern spiral 
arm, in a relatively low surface brightness region (Fig.1). This SN was
discovered by G. Valentini and coworkers (\cite{valent}) on Jan.26, 2000,
and immediately announced to be a Type Ia event by Turatto et al. 
(cf. IAUC~7351). They reported the appearance of 
\ion{Si}{ii}, \ion{Si}{iii}, \ion{S}{ii}, \ion{Ca}{ii} and \ion{Fe}{iii},
the usual ions characterizing Type Ia SNe, and also the
presence of Na~D indicating considerable reddening. The occurrence
of SN~2000E was particularly interesting, because it appeared just a
few months after the maximum of the Type II SN~1999el, the first 
SN observed in NGC~6951. SN~1999el was located closer to the
bar-dominated central region, but definitely outside the circumnuclear
regime where star-forming processes are most active (\cite{perez}). 

The expected maximum brightness of SNe Ia at the distance of NGC~6951
(about 13.5 mag) indicates that SN~2000E offers a good chance to increase
the sample of bright, well-observed SNe Ia. The comparison of
distances determined via SNe Ia and other methods may result in
the refinement of the distance measuring techniques and the cosmic 
distance scale itself. This is especially important in the case of
SNe, because they are used to measuring cosmological distances, where
other methods often do not work, and its technique relies on a relatively
small number of local calibrator SNe. 

In the followings the new photometric and spectroscopic observations of
SN~2000E are described then the results are presented and discussed. 

\section{Observations}

\begin{figure}
\leavevmode
\begin{center}
\psfig{file=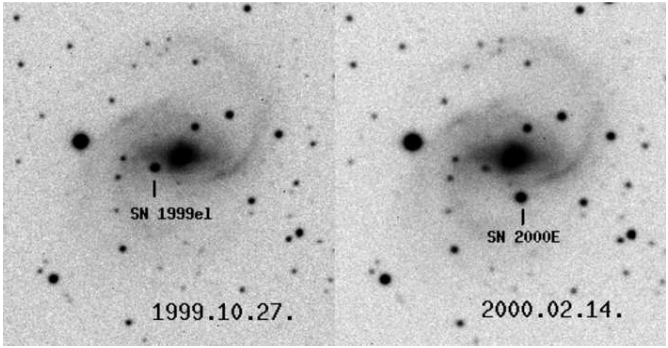,width=8.8cm}
\caption{NGC~6951 showing SN~1999el (left panel) and SN~2000E (right panel).}
\end{center}
\end{figure}

\subsection{Photometry}

The CCD-photometric observations were obtained with four telescopes:
the 28~cm Schmidt-Cassegrain at the campus site of University of
Szeged (\#1), the 60/90~cm Schmidt at Piszk\'estet\H o Station of
Konkoly Observatory (\#2), the 1~m Ritchey-Chr\'etien-Cassegrain at
Piszk\'estet\H o (\#3) and the 1.2~m Cassegrain at Calar Alto
Observatory, Spain (\#4). The small-aperture instrument was used only
around the maximum of the SN. The CCD-frames were exposed
through standard Johnson-Cousins filters, most often $V$ and $R_{\rm C}$.
$I_{\rm C}$ was used at early epochs, and a few $B$ frames
were also obtained at later phases. It would have been better to
use the same telescope and setup to obtain a homogeneous dataset.
However, this was strongly limited by the weather conditions and the 
availability of the instruments. 

Transformation to the standard system has been performed by applying
the following equations:

\begin{eqnarray}
V &=~ v + C_{\rm V} (V-R) + D_{\rm V}\\ 
R &=~ r + C_{\rm R} (V-R) + D_{\rm R}\\
I &=~ i + C_{\rm I} (V-I) + D_{\rm I}\\ \nonumber
\end{eqnarray}

The occasional $B$ data have been transformed in a similar way, except
that $(B-V)$ was used in the colour term.
The instrumental coefficients (the $C_i$ slopes and $D_i$ zero points)
have been determined by observing Landolt standard fields
(\cite{landolt}) or the M67 standard sequence (\cite{monty})
at each site. Actually, the use of the zero 
points have been eliminated by observing local comparison stars in the 
field of SN~2000E and computing differential magnitudes. Differential
extinction correction between the SN and the local comparison stars
was neglected, due to the small field of view. For each telescope 
the transformation slopes are collected in Table~1. The magnitudes of
the standard stars could be recovered within $\pm$ 0.03 magnitude.

\begin{table}
\caption{Transformation slopes. See text for the telescope codes.}
\begin{tabular}{cccc}
Tel. & $C_{\rm V}$ & $C_{\rm R}$ & $C_{\rm I}$ \\
  1  &$-$0.11&$-$0.08&$-$0.04\\
  2  &$+$0.10&$+$0.08&  -  \\
  3  &$-$0.06&$+$0.06&$+$0.05 \\
  4  &$+$0.06&$+$0.07&  -   \\
\end{tabular}
\end{table}

Fig.2 shows the field of NGC~6951 and SN~2000E with the local comparison
stars labelled. The standard magnitudes of these stars were determined
via Landolt standards observed with the Calar Alto telescope, where the 
photometric conditions were the best during our campaign. 
The results are listed in Table~2. Note that B1, B2 and B3 were used
only for the frames taken with telescope \#3.

\begin{table}
\caption{Standard magnitudes of local comparison stars. Errors are given
in parentheses.}
\begin{tabular}{ccccc}
Star & $V$   & $(B-V)$ & $(V-R)$ & $(V-I)$\\
 F1 &12.53 (0.01)&1.64 (0.02)&0.81 (0.01)&1.74 (0.01)\\
 F2 &14.86 (0.01)&0.87 (0.02)&0.52 (0.02)&1.01 (0.02)\\
 F3 &13.90 (0.03)&0.72 (0.04)&0.44 (0.01)&0.87 (0.02)\\
 F4 &15.55 (0.02)&0.74 (0.04)&0.43 (0.02)&0.92 (0.04)\\
 F5 &14.97 (0.02)&0.91 (0.03)&0.54 (0.02)&1.01 (0.02)\\
 F6 &15.71 (0.03)&0.88 (0.05)&0.49 (0.03)&1.07 (0.06)\\
 F7 &14.53 (0.02)&0.88 (0.02)&0.51 (0.02)&1.06 (0.03)\\
 F8 & saturated \\
 B1 &15.88 (0.03)&0.97 (0.05)&0.57 (0.03)&1.16 (0.04)\\
 B2 &16.42 (0.05)&1.01 (0.11)&0.54 (0.06)&1.21 (0.07)\\
 B3 &16.60 (0.05)&1.18 (0.11)&0.64 (0.06)&1.32 (0.07)\\ 
\end{tabular}
\end{table}

The magnitudes of SN~2000E were inferred with aperture photometry, and 
transformed to the standard system via the local comparison stars.
A small aperture radius of 4 pixels has been used in order to
minimize the effect of the host galaxy background. The background light
was estimated within a 4 pixel-wide annulus having 5 pixels inner radius. 
This background was determined on each frame and 
subtracted from the SN flux. PSF-photometry was not applied, because the frames 
with telescopes \#1 and \#2 had undersampled and/or strongly varying PSF. 
The transformation equations (Eq.1-3) were applied for all observed SN
data including those that were obtained more than 1 month past maximum.
The colour terms in Eq.1-3 resulted in magnitude corrections in the order 
of 0.05 mag in the nebular phase. It may caused some additional uncertainty,
because at late epochs the spectral distribution of the SN light 
resembles more a nebula than a star. K-correction has been neglected, 
because at the redshift of NGC~6951 ($z = 0.005$) it does not exceed 
0.01-0.02 mag in $V$ (\cite{hamuk}), i.e. it is well below the
photometric uncertainty. 
 
The effect of the host galaxy background on the SN flux was investigated 
on frames taken in October, 1999 showing SN~1999el only 
(see Fig.1, left panel). An aperture with the same size was
placed on the position of SN~2000E and the surrounding background
was subtracted, as above. The remaining flux was negligible even in the
$R$ filter. Thus, the background subtraction with the small aperture-annulus
combination gave acceptable magnitudes for SN~2000E, minimizing the
effect of the host galaxy light. 

\begin{figure}
\leavevmode
\begin{center}
\psfig{file=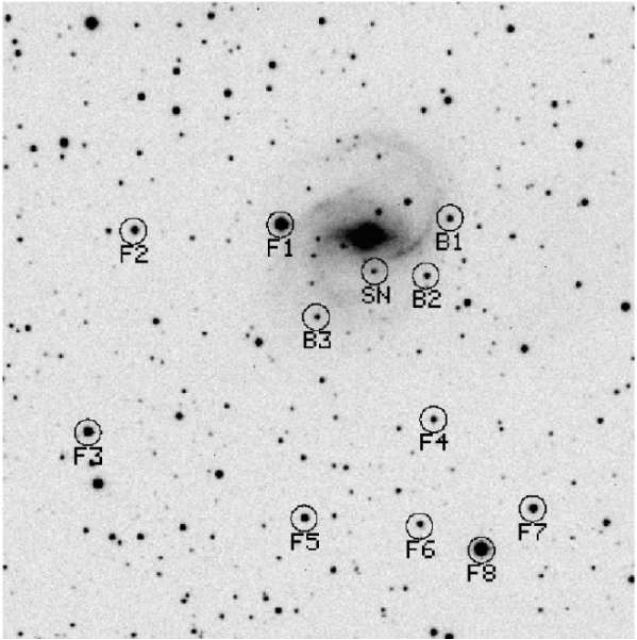,width=8.5cm}
\caption{Local secondary standard stars in the field of NGC~6951. 
North is up and east is to the left. See Table~2 for magnitudes.}
\end{center}
\end{figure}
 
The list of standard magnitudes of SN~2000E is given in Table~3.
The errors were estimated from the rms deviations of the SN
magnitudes from the comparison stars on each frame, and do not
contain the possible systematic errors arise from the standard
transformation. We believe that this latter error source is smaller
than the uncertainites of the instrumental magnitudes measured
with smaller telescopes.

The magnitudes in Table~3 were also compared with  
filtered CCD-magnitudes of \cite{hh} that were made
between JD~51655 and 51666. Their $R$-magnitudes differ
by about 0.4 mag from our values. This is a 4$\sigma$
difference considering the given uncertainty 
of their data ($\pm$0.1 mag). Because the details of
the standard transformation of the data of Hornoch \& Hanzl
were not published, the cause of this discrepancy
cannot be studied in more detail here, except to underline
that this may indicate a systematic error in either
datasets at late epochs of SN~2000E. More published
standardized measurements are needed to resolve this
issue.

\begin{table*}
\caption{Photometric data of SN~2000E. Errors are given in parentheses.}
\begin{tabular}{cccccc}
JD & $B$ & $V$ & $R_{\rm C}$ & $I_{\rm C}$ & Tel.\\
2451572.2 & -  &  -  & 13.77 (0.03)&   -  & 2\\
2451576.2 & -  &13.88 (0.07)&13.53 (0.04)&13.72 (0.06)& 1\\
2451578.3 & -  &13.73 (0.04)&13.51 (0.02)&13.57 (0.06)& 1\\
2451579.3 & -  &13.67 (0.04)&13.51 (0.08)&13.68 (0.05)& 1\\
2451581.3 & -  &13.78 (0.10)&13.50 (0.16)&13.59 (0.15)& 1\\
2451582.3 & -  &13.80 (0.11)&13.63 (0.08)&13.60 (0.05)& 1\\
2451585.3 & -  &14.02 (0.07)&13.76 (0.04)&13.75 (0.05)& 1\\
2451589.5 & -  &14.21 (0.02)&14.12 (0.02)&14.14 (0.03)& 2\\
2451656.5 & -  & -   &16.39 (0.02)&   -  & 2\\
2451657.5 & -  & -   &16.43 (0.02)&   -  & 2\\
2451658.5 &17.64 (0.02)&16.65 (0.03)&16.48 (0.02)&  -  & 2\\  
2451661.5 &17.66 (0.02)&16.79 (0.02)&16.52 (0.02)&  -  & 2\\
2451662.5 &17.80 (0.03)&16.88 (0.02)&   - &  -  & 2\\
2451664.5 &17.85 (0.02)&16.93 (0.02)&   - &  -  & 2\\
2451667.5 & -  &16.82 (0.10)&16.51 (0.15)&   -  & 2\\
2451696.5 & -  &17.63 (0.02)&17.24 (0.03)&   -  & 3\\
2451705.6 & -  &17.89 (0.04)&17.87 (0.08)&18.09 (0.08)& 3\\
2451706.6 & -  &17.89 (0.06)&17.86 (0.10)&  -   & 3\\
2451727.4 & -  &18.22 (0.02)&18.41 (0.05)&18.24 (0.05)& 4\\
\end{tabular}
\end{table*}

The light curves in $V$, $R$ and $I$ filters are plotted in
Fig.3. The continuous line is the optimal result of the
Multi-Colour Light Curve Shape (MLCS) method (see next section).
It can be revealed that the maximum brightness in $V$ occurred
around JD 51580, and the peak $V$-magnitude was about 13.7 mag,
being in good agreement with the predicted peak brightness of
SNe Ia at the distance of NGC~6951 (see Sect.1).

\begin{figure}
\leavevmode
\begin{center}
\psfig{file=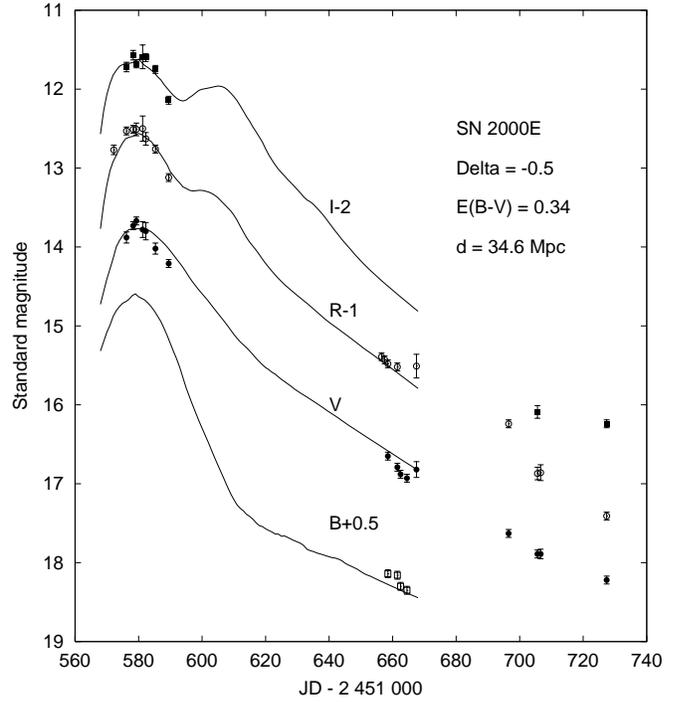,width=9cm}
\caption{Light curves of SN~2000E in $B,V,R,I$ filters. The $B,R,I$ data
have been shifted vertically for better visibility. The
continuous line is the template calculated with the MLCS method
using $\Delta=-0.5$, $E(B-V)=0.34$ and $\mu_{\rm 0}=32.70$ mag (see Table 4).}
\end{center}
\end{figure}

\begin{figure}
\leavevmode
\begin{center}
\psfig{file=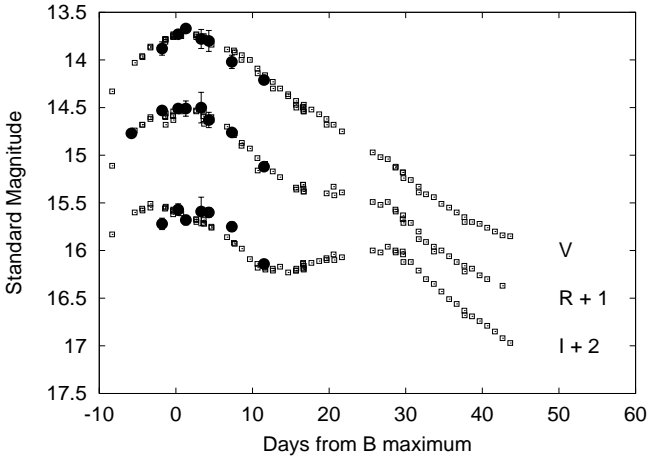,width=9cm}
\caption{Comparison of $V(RI)_{\rm C}$ light curves of SN~2000E 
 (filled circles) with those of SN~1998bu (open symbols). 
 All SN~1998bu data have been shifted vertically by 1.88 mag
 in order to bring them into agreement with SN~2000E. 
 The similarity of the light curves is apparent.}
\end{center}
\end{figure}

In order to test the reliability of our measurements
and standard transformation, the light curves of SN~2000E were
compared with those of SN~1998bu. 
For this bright SN, which appeared
in the Leo I group galaxy M96, published light curves
of very good quality are available, 
and it received considerable attention
recently (\cite{jha}; \cite{sun}). Fig.4 shows that the
$V$, $R$, $I$ magnitudes of SN~2000E (filled symbols) 
at the early part of
the light curve agree satisfactorily with the data of
SN~1998bu. This agreement was reached by 
adding 1.88 mag to the light curves of SN~1998bu in all filters. 
This suggests that the reddening of the two
SNe were similar, about $E(B-V) \approx 0.3$ mag 
(\cite{jha}; see also Sect.3). 
At present, the accuracy of inhomogeneous SNe light curves 
is usually not better than $\pm$0.1 (see e.g. Fig.8
of \cite{jha}), so the small deviations in Fig.4 (especially
in the $I$ band) are probably not significant.
The light curves will be analysed further in Sect.3.

\subsection{Objective-prism spectroscopy}

Spectroscopic observations were gathered with an objective 
prism attached to the 60/90 cm Schmidt telescope at 
Piszk\'estet\H o Station 
of Konkoly Observatory, between 26th and 28th February, 2000 
(JD 2451601 - 03), when SN~2000E was about 1 month past maximum.
The images were exposed onto an electronically cooled
Thomson $1536 \times 1024$ CCD-chip (readout noise about 16 $e^-$).
The dispersion axis was aligned in the north-south direction, along
the shorter side of the CCD-chip.

An objective prism spectrograph is certainly not an ideal tool for
SN spectroscopy. However, this was the only spectroscopic 
instrument available to us at that time. 

A considerable number of image processing steps were necessary
to extract the SN from the smeared spectrum of the host galaxy.
The location of the SN spectrum was determined from an intensity
plot along the line perpendicular to the dispersion axis.
At first, this was possible only in the blue region where the
host galaxy showed negligible contribution. The red side, however,
was heavily contaminated by the smeared background spectrum of
the host galaxy. Removal of this background was essential
to obtain reliable SN spectra.

The easier way to correct for the galaxy background would have
been the usage of an objective prism picture of the host
galaxy without the SN. Since it was not possible for us, we
had to choose another, more approximative approach. First, 
the central peak of the background galaxy spectrum was identified
visually. Then, the galaxy image was cut into two pieces
at this central ridge, and the western part (including the
SN) was dropped. The eastern side was reflected and
added back into the position of the western side, thus,
generating a symmetric picture of the galaxy. This picture
was then subtracted from the original one, resulting in
a much cleaner SN spectrum. While the removal of the 
galaxy ``spectrum'' was far from complete, its contribution
at the position of the SN spectrum was considerably suppressed in
this way.

\begin{figure}
\leavevmode
\begin{center}
\psfig{file=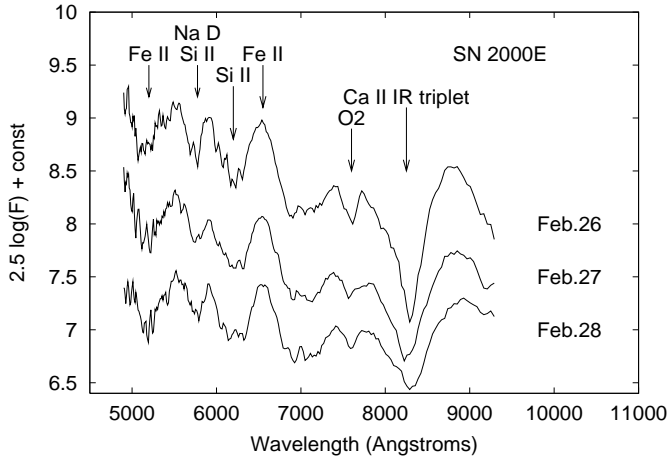,width=9cm}
\caption{Objective-prism spectra of SN~2000E. The epochs of the 
observations are indicated to the right of each spectrum. Some
spectral features are marked (see text for reference).}
\end{center}
\end{figure}

The extraction of the cleaned SN spectrum was performed
with the standard subroutines in 
{\it IRAF/SPECRED}\footnote[1]{{\it IRAF} is distributed by NOAO
which is operated by the Association of Universities for Research in Astronomy 
(AURA) Inc. under cooperative agreement with the National Science Foundation.}.
The intensities within a 3-4 pixel-wide aperture were
summed and this aperture was slid along the dispersion
axis, taking into account the tilt of the spectrum.
The remaining background light was subtracted after a
low-order polynomial fit. The wavelength calibration
was performed using the lines in the spectrum of Vega 
taken with the same instrument and setup. The Vega spectrum
was also used for the flux calibration. First, the telescope
response function was determined by matching the measured 
continuum fluxes of the Vega spectrum with the tabulated ones
(given in e.g. \cite{gray}). Then, the SN spectrum was multiplied
by the response function producing a flux-calibrated spectrum.
The resulting spectra of SN~2000E are plotted together 
in Fig.5, where an arbitrary vertical shift was applied 
for better visibility. Information for line identification
was collected from \cite{filip1}.

It is apparent from Fig.5 that SN~2000E shows the standard
spectral features of a Type Ia SN at one month after maximum
light. A closer inspection of the features around 6000 \AA\ 
with those of SN~1998aq taken at the same phase (\cite{vinko})
showed good agreement, despite of the much lower resolution of
the present spectra. This confirms the classification of
Type Ia, although a slightly confusing description of the
presence of the $H\alpha$ line was also reported in IAUC~7353 by
\cite{polc}. This was then revised in IAUC~7359. 

There is continuously growing amount of evidence that 
SNe~Ia show considerable diversity in peak brightness, 
decline rate, spectral features, etc. 
(see e.g. \cite{filip1, phil, nugent, hat} and references therein).
The similarity between the spectra of SN~2000E and SN~1998aq
may mean that SN~2000E is close to the ``normal'' SNe Ia being
neither SN~1991T-like, nor SN~1991bg-like event. 
It should be noted, however, that the spectroscopic
diversity between these subclasses of SNe Ia is usually studied
at earlier epochs, between $\pm 10$ days around maximum. Therefore,
the spectra presented here are too late for such a distinction.
This issue will be studied in more detail using the shape of the 
light curve in the next section.

\section{Results and discussion}

In this section, first, we estimate the reddening of SN~2000E,
then its distance is determined from the analysis of the light
curve. Finally, the discussion of the errors of distance 
measurement is presented.

\subsection{Reddening}
The presence of Na~D was reported in the very first spectrum
of SN~2000E (see Sect.1) indicating significant 
reddening.  This feature might also be visible in the 
spectra in Fig.5, although it is quite difficult to identify
the heavily blended lines at the end of the photospheric phase.
Nevertheless, the determination of reddening is essential 
if SN~2000E is to be used for distance measurement.

The reddening map of \cite{buhe} indicates $E(B-V)=0.20$ at
the position of SN~2000E, while the more recent map of
\cite{sfd} gives 0.36 mag. This is referred as the
galactic component of the reddening, originating mostly from
the interstellar medium (ISM) within the Milky Way.
The map of Schlegel et al. is thought to be a better
representative of the true amount of reddening, but it has been
pointed out that this map systematically overestimates the
reddening at directions where $E(B-V) \geq 0.15$ mag 
(\cite{arce}). Since several galactic cirrus clouds are visible 
on the long-exposure frames of NGC~6951, substantial reddening
due to the Milky Way ISM is expected. Taking into account
the possible overestimate in the map of Schlegel et al., the
galactic component of $E(B-V)$ may be somewhere between the
two values given above (0.2 - 0.36 mag). 

The total reddening of SN~2000E was estimated by comparing
the observed $(B-V)$ index with its expected value at the
given epoch. This may give reasonable results, because
SNe Ia are thought to show some kind of homogeneity 
in their $(B-V)$ colours at epochs later than 60 
days post-maximum (\cite{phil}).  
Using JD~51578 as the epoch of the $B$-maximum (see next 
section), the phase of the $B$-data in Table~3 is between
$\tau$ = 80 and 86 days. The empirical formula of \cite{phil} 
results in $(B-V)_0 = 0.49 ~\pm~ 0.05$ 
at $\tau$ = 80 days, while the tabulated $(B-V)_0$ 
of \cite{riess1} gives 0.73 $\pm$ 0.02 at this epoch. 
It is apparent, that, although both values are based on 
empirical data of ``standard'' SNe Ia, their difference 
is large and significant. Certainly, this method of 
reddening determination has ambiguity, and the result 
may strongly depend on the adopted fiducial $(B-V)_0$. 
The referee of this paper, Kevin Krisciunas, mentioned 
that his own analysis of recent SNe (\cite{kris}) 
gave a $(B-V)_0 ~-~ \tau$ 
relation with a zero point 0.070 mag redder than in \cite{phil}.

Nevertheless, as a first approximation, 
we decided to adopt an averaged $E(B-V)$ using the values above. 
Thus, the reddening was
calculated by subtracting the expected $(B-V)_0$s from
the observed ones at the epochs when the $B-V$ 
observations were taken, and then averaging
the results. The Phillips et al.-relation resulted in 
$E(B-V)=0.49 \pm  0.02$ mag, while the Riess et al.-method
produced $E(B-V)=0.24 \pm 0.01$ mag. Their average,
$E(B-V)=0.36 \pm 0.15$ was accepted as the best result at
present. This is in very good agreement with the value
of \cite{sfd}. It suggests that most of the observed reddening
of SN~2000E is due to dust in the Milky Way. 
The rather large uncertainty ($\pm$ 0.15 mag) reflects
the ambiguity of choosing the intrinsic $(B-V)_0$ of
SNe Ia at late epochs. Using the standard $R_{\rm V} = 3.1$ 
coefficient of the galactic reddening law, the resulting
uncertainty of the total absorption is 
$\Delta A_{\rm V} = \pm 0.46$ mag.

\subsection{MLCS method}

The light curves of Type Ia SNe correlate with their peak brightness
such that intrinsically brighter SNe Ia are bluer at maximum
and decline more slowly than intrinsically dimmer SNe. 
The Multi-colour Light Curve Shape ({\it MLCS}) method parametrizes 
the light curve family by introducing $\Delta$ that measures approximately 
the $V$ magnitude difference of a particular SN from a fiducial light curve 
at the time of $B$-maximum. The first version of the MLCS-method (MLCS-1, 
\cite{riess1}) assumed a linear dependence of the light curve shape 
on $\Delta$, which was calibrated via a few nearby SNe. The second version 
(MLCS-2) uses a quadratic dependence on $\Delta$, and it is based
on a more extended set of SNe light curves (\cite{riess2}).  

The light curves of SN~2000E (Table~3) were analysed with the 
second-order MLCS-2 method. The observed magnitude of the SN
in filter $k$ ($=B,V,R,I$) at a particular epoch $\tau$ was described as
$$ m_k(\tau) ~=~ M^{max}_k + M_k(\tau) + R_k(\tau) \Delta + Q_k(\tau) 
\Delta^2 + \mu_{\rm 0} + A_k $$
where $M^{max}_k$ is the magnitude zero point at the time of $B$-maximum 
($\tau = 0$), $\mu_{\rm 0}$ is the true distance modulus, $A_k$ is
the total extinction in the given filter, and $M_k(\tau)$, 
$R_k(\tau)$, $Q_k(\tau)$ are the template vectors of
MLCS-2. The template vectors were kindly supplied to us by
the referee with the permission of Adam Riess. The $M^{max}$
fiducial zero points were chosen as $-$19.46, $-$19.46, $-$19.41 and
$-$19.81 mag for $B,V,R,I$ filters, respectively. 
The interstellar extinction was calculated by
assuming the galactic reddening law with coefficients given by
\cite{sfd}: $A_{\rm V} = 3.1 E(B-V)$, $A_{\rm B} = 4.1 E(B-V)$,
$A_{\rm R} = 2.46 E(B-V)$, $A_{\rm I} = 1.72 E(B-V)$. Thus, the observed light
curve is modelled via four free parameters: the time of $B$-maximum 
$T_{\rm max}(B)$, the true distance modulus $\mu_{\rm 0}$, the reddening $E(B-V)$ and
the luminosity parameter $\Delta$. 

The template vectors were fitted to the light curves of SN~2000E
simultaneously, i.e. the difference between the observed and
calculated light curves were combined into a single chi-squared
function. The weighting factors were chosen as $1/\sigma_i^2$
where $\sigma_i$ is the photometric uncertainty of the given data point
(listed in Table~3). In order to avoid giving too strong weight to any specific
data point, the minimal photometric 
uncertainty was increased to 0.05 mag. Because the template vectors 
are better determined at early epochs than at the nebular phase,
more weight was assigned to the data around maximum
(Riess, personal communication). For this purpose, 
the error bars at late epochs ($\tau >$70 days)
were multiplied by $\sqrt 2$ thus producing a factor of
2 less weight for these data.     

Three kinds of solutions have been determined with the MLCS-2 vectors.
First, $E(B-V) = 0.36$, estimated in Sect.3.1, was treated fixed and
only $\mu_{\rm 0}$ and $\Delta$ were optimized. The solution converged
to $\mu_{\rm 0} = 32.62 \pm 0.1$ and $\Delta = -0.5 \pm 0.1$. 
Second, relaxing this constraint
and optimizing all three parameters we have got $E(B-V) = 0.34 \pm 0.05$,
$\mu_{\rm 0} = 32.70 \pm 0.1$ and $\Delta = -0.5 \pm 0.05$. These two
solutions agree quite well and suggest that
this SN belongs to the overluminous subclass of SNe Ia. Finally,
forcing the solution only to the data around maximum (dropping
the points at late epochs) resulted in $E(B-V) = 0.32 \pm 0.08$,
$\mu_{\rm 0} = 32.12 \pm 0.2$ and $\Delta = +0.14 \pm 0.1$. Thus, the
data around maximum favour slightly positive $\Delta$ and
a shorter distance, while late-time photometry, which gives
tighter constraint on $\Delta$, indicates strongly negative
$\Delta$ and a higher distance. The two solutions are significantly   
different from each other, the distance moduli of solution \#2 and \#3 
differs by $3 \sigma$. Of course, $\mu_{\rm 0}$ and $\Delta$ are strongly correlated 
parameters, a negative $\Delta$ (brighter SN) results in a larger 
distance modulus for the same $E(B-V)$.
Note that the application of MLCS-1 
(the previous, linear version of the method), 
to the whole light curve resulted in $E(B-V) = 0.30 \pm 0.1$, 
$\mu_{\rm 0} = 32.16 \pm 0.1$ and $\Delta = +0.15 \pm 0.1$, which is
closer to the MLCS-2 solution \#3. This is due to the fact 
that the MLCS-1 fiducials produce brighter light curves at 
late phases than the MLCS-2 vectors, so a smaller $\Delta$ is 
needed to fit the late data of SN~2000E. 
The parameters of these solutions are collected in Table~4.

\begin{table}
\caption{Parameters of SN~2000E optimized with MLCS. 
$T_{\rm max}(B)=51578.0$ was used in all solutions. Uncertainties are given
in parentheses. \#1-3 were computed with MLCS-2, while \#4 was
obtained with MLCS-1.}
\begin{tabular}{lcccc}
No.&$E(B-V)$&$\mu_{\rm 0}$&$\Delta$&$\chi^2$\\
\#1&0.36 (fixed)&32.62 (0.1)&$-$0.50 (0.10)&2.32\\
\#2&0.34 (0.05)&32.70 (0.1)&$-$0.50 (0.05)&2.27\\
\#3&0.32 (0.08)&32.12 (0.2)&$+$0.14 (0.10)&6.58\\
\#4&0.30 (0.10)&32.16 (0.1)&$+$0.15 (0.05)&2.52\\
\end{tabular}
\end{table} 

Is it possible that SN~2000E had $\Delta = -0.5$? Such overluminous
SNe Ia often belong to the ``SN~1991T'' subgroup, which show
peculiar premaximum spectra: almost featureless continuum with 
no sign of Si or S, but a few ionized Fe lines (Filippenko, 1997). 
According to the description given in Sect.1, Si~II, S~II and
Ca~II could be identified in the premaximum spectra of SN~2000E,
which argues against an SN~1991T-like object. 
The postmaximum spectra of such SNe are more-or-less normal,
so the spectra of SN~2000E presented in Sect.2.2 cannot be used
for identifying this subclass. On the other hand, there is an
example of a strongly overluminous SN ($\Delta \approx -0.5$),
SN~1992bc, that
otherwise had normal premaximum spectrum 
(\cite{hamum}; \cite{riess2}). 
If the $\Delta = -0.5$ solution of MLCS-2 is true then 
SN~2000E might be similar to SN~1992bc.
    
How can we get closer to the ``true'' value of $\Delta$? 
$\Delta$ is constrained by both the shape of the light curve and
the colour at maximum. SNe with $\Delta < -0.1$ show the bump in their 
$I$ light curve at later epochs than SNe with $\Delta \approx 0$ do.
Unfortunately, the observed $I$ curve of SN~2000E does not
extend into the bump phase, so this constraint could not
be applied. Alternative methods, such as the $\Delta m_{\rm 15}(B)$ 
decline rate of the early $B$ light curve (\cite{phil0}; 
\cite{hamuy0}), or infrared $JHK$
photometric templates (\cite{kris2}) could not be used for
the same reason, i.e. the lack of the necessary data. The colour
around maximum, in principle, may be indicative of a negative $\Delta$,
because overluminous SNe are bluer at maximum. However, this is
distorted by reddening which must be determined separately to use
this constraint. Uncertainty in the reddening would directly influence
the inferred $\Delta$, especially in the $\Delta < 0$ domain, because
a negative $\Delta$ corresponds to less colour change than a positive
$\Delta$ in MLCS-2. 

The fact that the early light curve of SN~2000E 
was very similar to that of a normal SN Ia makes the determination
of $\Delta$ very difficult from the present dataset. The result
of $\Delta = -0.5$ greatly relies on our late-epoch photometry.
Because at late epochs the light of the SN is more affected
by the background of its host galaxy, an incorrect background
subtraction could easily lead to brighter SN magnitudes, thus,
suggesting a slower decline and a negative $\Delta$. We have
checked the efficiency of the background subtraction (Sect.2.1),
still, such systematic error cannot be ruled out completely.  
Also, the standard transformation is more uncertain in the nebular
phase, as mentioned in Sect.2. It is not possible to reach an
unambiguous result at present. Thus, we conclude that the available
data suggest that SN~2000E was overluminous at maximum, but otherwise
showed normal spectral features, similarly to SN~1992bc.

\subsection{The distance of NGC~6951}

The MLCS analysis described above resulted in essentially two sets of
distances to the host galaxy of SN~2000E. The ``long'' distance,
corresponding to $\Delta = -0.5$, is $d$ = 34 $\pm$ 2 Mpc 
(this uncertainity reflects simply the difference between the two
solutions in Table~4), while the ``short'' distance is about
$d$ = 26 $\pm$ 3 Mpc.
The $d$ = 34 Mpc distance is more likely, but the ``short'' distance
(meaning that SN~2000E was close to the fiducial SN Ia) cannot
be ruled out. 

The MLCS distances are tied to a Cepheid distance scale (\cite{riess1};
\cite{jha}, and references therein). This can be checked by comparing
the light curves and distance moduli of SN~2000E and SN~1998bu.
As presented in Sect.2.1, the light curves of these two SNe differs
by 1.88 mag uniformly in all $VRI$ bands. Rearranging the basic equation
of the MLCS method, around maximum in $V$ filter one can get
$$\mu_{\rm 0}(E)=\mu_{\rm 0}(bu)+(m(E)-m(bu))-$$
$$~~~~(A_{\rm V}(E)-A_{\rm V}(bu)) - R_{\rm v} (\Delta(E) - \Delta(bu)) \eqno (5) $$
where $E$ denotes SN~2000E and $bu$ is for SN~1998bu and
all symbols have their usual meaning (the $Q \Delta^2$ term
is negligible around maximum). The terms on the right-hand side
are $\mu_{\rm 0}(bu) = 30.37$ mag (\cite{jha}), $m(E) - m(bu) = 1.88$ mag
(Sect.2.1), $A_{\rm V}(E) - A_{\rm V}(bu) = 0.15$ mag (adopting $E(B-V) = 0.31$
for SN~1998bu, \cite{jha}), $R_{\rm v} \approx 1$ and 
$\Delta(E) - \Delta(bu) = -0.52$. These values give 
$\mu_{\rm 0}(E) = 32.62$ mag, in good agreement with the MLCS result (Table~4).
Similarly, the solutions \#3 and \#4 in Table~4 would give 
$\mu_{\rm 0}(E) = 32.08$ mag which agrees with the MLCS distance modulus
within the error. The small difference is attributed to the ambiguity
of the Cepheid distance scale. For example, \cite{key1} derived
$\mu_{\rm 0} = 30.20 \pm 0.1$ mag for the Cepheid distance modulus of M96,
which differs by 0.17 mag from the value used by \cite{jha}. 
Using this for $\mu_{\rm 0}(bu)$ one could get systematically less
distance modulus for SN~2000E. Accordingly, the SN-distances
are uncertain by about 0.2 mag due to the uncertainties of
the adopted distance scale. The discussion of the distance scale
calibrations is beyond the scope of this paper. 

The distance to NGC~6951 was finally calculated by adopting the 
average of the ``long'' distance moduli (32.70 by MLCS,
32.62 by the Jha et al.-distance to M96 and 32.45 by the 
Gibson et al.-distance to M96). The result is 
$\mu_{\rm 0} = 32.59 \pm 0.1$ mag, or $33 \pm 2$ Mpc. 
Note that the similar average of the ``short'' 
distances results in 32.28 $\pm$ 0.2 mag, 
or 29 $\pm$ 3 Mpc.
  
The new distance of NGC~6951 is significantly larger 
(by about 0.7 mag, or 9 Mpc) than the previous Tully-Fisher
distance estimates (see Sect.1). This is close to the
``usual'' systematic difference between the two distance
scales (e.g. \cite{riess1}). Recently \cite{shanks} 
proposed a modification of Tully-Fisher distances 
by adding 0.46 mag to T-F distance moduli to bring 
them into agreement with SNe Ia distances.
This correction is less than the difference above, but it is
known that the T-F distance moduli of individual galaxies
can be uncertain by at least 0.3 - 0.4 mag. Naturally,
the ``short'' distance given above would agree better
with the T-F distances than the ``long'' distance. 
Note that a recent revision of the Cepheid distance scale
by \cite{key1}, suggests a better agreement with the 
Cepheid and T-F distance scales. Indeed, the distance 
of SN~2000E would agree with the previous T-F distance 
of NGC~6951 by \cite{tully}, if the SN were normal 
(i.e. not overluminous) and the distance scale of
\cite{key1} was used.

The referee suggested a check of the inferred distance of
NGC~6951 based on the measured redshift and the expected Hubble-flow
at that distance. The SIMBAD catalogue gives 
$v_{\rm rad}$(NGC~6951) = 1424 kms$^{-1}$. The galactic coordinates
of this galaxy are $l = 100^{\circ}.89$ and $b = 14^{\circ}.85$, thus,
the radial velocity in the galactic system is represented
by a vector of (+365; +1352; $-$260) kms$^{-1}$. 
The additive correction for the Milky Way motion within the 
Local Group can be computed as ($-$30; +297; $-$27) (\cite{riess1}).
The drift of the Local Group with respect to the CMB can be
approximated by adding (+10; $-$542; +300) (\cite{smoo}) or
(+57; $-$540; 313) (\cite{kogut}). Both corrections results in
$v_{\rm rad}$(NGC~6951) = 1307 kms$^{-1}$ in the
CMB rest frame. The expected Hubble-flow velocity at the inferred 
distance of NGC~6951 ($d = 33$ Mpc) is 2145 kms$^{-1}$ 
using $H_{\rm 0} = 65$ kms$^{-1}$Mpc$^{-1}$.
This is significantly larger than the corrected radial
velocity of NGC~6951. However, it is not an unexpected result,
because at this distance the observed radial velocities usually
deviate from the smooth Hubble-flow. Note that using a
recent estimate of the short-distance scale Hubble-constant
$H_{\rm 0} = 76$ (\cite{jensen}) one would get $v_{\rm rad} = 2508$ kms$^{-1}$,
increasing further the discrepancy. The shorter SN-distance
(29 Mpc) with $H_{\rm 0} = 65$ would lead to $v_{\rm rad} = 1885$ kms$^{-1}$,
which is still higher than the observed rest-frame 
radial velocity. So, the conclusion is that
NGC~6951 is probably too close to predict a reliable radial 
velocity from the smooth Hubble-flow.  

\subsection{Error budget}

The overall error of the distance moduli determined above
contains contribution from measurement errors (random and
systematic) and the systematic uncertainties of 
several assumptions. 

The standard deviation of the MLCS distance can be expressed
as
$$
\sigma^2_{\mu_{\rm 0}} = \sigma^2_m + \sigma^2_M + \sigma^2_{A_{\rm V}} + \sigma^2_R
$$
where $\sigma_m$ and $\sigma_M$ are the uncertainties of
the measured and the template $V$ light curve, 
$\sigma^2_{A_{\rm V}} = 3.1^2 \sigma^2_{E(B-V)}$ and
$\sigma_R$ is the uncertainty of the $R_{\rm v} \Delta$ product
(see \cite{riess1}). Assigning $\sigma_m = 0.05$ mag (random), 
$\sigma_M = 0.15$ mag (systematic zero-point error
and random scattering of the template light curves), 
$\sigma_{E(B-V)} = 0.15$ mag (random measurement error and
systematic uncertainty of the fiducial $(B-V)_0$) 
and $\sigma_R = 0.1$ (systematic) the result is 
$\sigma_{\mu_{\rm 0}} = 0.50 ~{\rm mag}.$ 
Note that this uncertainty
does not contain the possible systematic error of the
standard transformation of the SN-photometry.

The Cepheid-based distance via SN~1998bu has lower
uncertainty, because in this case the errors of the 
parameter {\it differences} are considered. 
From Eq.4 the error can be expressed as
$$
\sigma^2_{\mu_{\rm 0}}(E) = \sigma^2_{\mu_{\rm 0}}(bu) + \sigma^2_{\Delta m}
+ \sigma^2_{\Delta A} + {R_v}^2 (\sigma^2_{\Delta}(E) + \sigma^2_{\Delta}(bu))
$$
where $\sigma_{\Delta m}$ is the error of the magnitude shift
between the two sets of light curves, $\sigma_{\Delta A}$ is 
the uncertainty of the $A_{\rm V}(E) - A_{\rm V}(bu)$ difference. 
These uncertainties are estimated as $\sigma_{\Delta m} = 0.03$, 
$\sigma_{\Delta A} = 0.1$, $\sigma_{\Delta}(bu) = 0.02$
(Sect.3.3), $\sigma_{\Delta}(E) = 0.1$ (as above), and
$R_{\rm v} \approx 1$ around maximum. The error of the distance modulus of
SN~1998bu was considered to be $\sigma_{\mu_{\rm 0}}(bu) = 0.2$ mag
taking into account both random and systematic uncertainties
(see Sect.3.3). Substituting these into the above expression,
the result is  $\sigma_{\mu_{\rm 0}} = 0.25 ~{\rm mag}.$

The uncertainties above do not contain the possible large
systematic error of the luminosity parameter $\Delta$
(i.e. the problem of the ``long'' vs. ``short'' distances
of SN~2000E). Clearly, $\Delta$ has a crucial role in
the distance determination, therefore, any systematic error
in $\Delta$ directly distorts the photometric distance.
If we adopt a rather pessimistic viewpoint, and assign 0.5 mag to 
the systematic error of $\Delta$ (which would mean that the
slower decline of SN~2000E is due to systematic errors
in our photometry) the error of the MLCS distance modulus 
grows up to $\sigma_{\mu_{\rm 0}} = 0.71$ mag, 
while the distance modulus via SN~1998bu  has 
uncertainty of $\sigma_{\mu_{\rm 0}} = 0.56$ mag.

The other factor that contributes significantly to
the total uncertainty is the reddening $E(B-V)$.
It is multiplied by 3.1 
to take into account the total absorption, which,
unfortunately, greatly amplifies the overall uncertainty.
Generally, SNe Ia are better distance indicators than
the case presented here, because most of them have
much lower reddening. However, it cannot be expected
that $all$ SNe Ia have negligible reddening, thus,
the uncertainties in $E(B-V)$ can seriously disturb
even the distances of relatively nearby SNe. Because
the better understanding of SNe Ia greatly relies on
the nearby, well-observed events, it is essential to
work out reliable methods for the reddening determination.
The increasing number of such objects would certainly help 
to solve this problem.

\section{Summary}

\begin{enumerate}
\item{$V(RI)_{\rm C}$ photometry of SN~2000E in NGC~6951 was obtained
starting from 1 week before maximum light and extending up
to 150 days past maximum. The shape of the light curves and
the objective-prism spectra collected at 1 month past maximum
confirm the Type Ia nature of SN~2000E.}

\item{The reddening of the supernova is estimated from the
comparison of the measured and expected $(B-V)$ colour at
late epochs (around 80 days past maximum). Two different
template $(B-V)$ curves were used resulting in 
$E(B-V) = 0.36 \pm 0.15$ as an averaged final result.
It is consistent with the prediction of the galactic 
reddening map of \cite{sfd}. The error reflects 
the systematic uncertainties of the template colour curves
at late epochs.}

\item{The distance of SN~2000E has been inferred from
the second-order MLCS-method. The fitting to $BV(RI)_{\rm C}$
light curves resulted in $\mu_{\rm 0} = 32.70 \pm 0.50$ and
$\Delta = -0.5 \pm 0.1$ mag as true distance modulus
and light curve parameter, respectively. The negative 
$\Delta$ suggests that SN~2000E was overluminous relative
to the fiducial Type Ia supernova, but this is based on
inhomogeneous, low signal-to-noise photometric data, thus, it may be 
systematically in error.}

\item{The similarity of the light curve of 
SN~2000E with that of SN~1998bu in M96 allowed us to tie the
distance of SN~2000E to the $HST$ Cepheid distance scale.
$\mu_{\rm 0} = 32.62$ - $32.45 ~\pm 0.26$ mag was obtained 
depending on the adopted Cepheid-distance modulus of
M96, which agrees with the MLCS-distance given above.
Averaging the distance moduli of SN~2000E supplied by 
various methods, the estimated distance modulus becomes 
$$ \mu_{\rm 0} = 32.59 \pm 0.5 ~{\rm mag}$$
corresponding to $33 \pm 8$ Mpc geometric distance.
The largest error sources in the distance estimates 
are the systematic errors in $\Delta$ and $E(B-V)$.
If the $\Delta$ parameter of SN~2000E were determined 
incorrectly, the distance would be systematically
lower and its uncertainty would grow up to $\pm$ 10 Mpc.}

\end{enumerate}

\begin{acknowledgements}

This work was supported by Hungarian OTKA Grants
No. T032258, T034615, the Magyary Postdoctoral Fellowship to JV from 
Foundation for Hungarian Education and Science (AMFK), 
the ``Bolyai J\'anos'' Research Scholarship to 
LLK from Hungarian Academy of Sciences
and Pro Renovanda Cultura Hungariae Foundation (travel 
grants to BCs, KS and GySz). We are grateful to the referee, 
Dr. Kevin Krisciunas, who made several contributions 
that helped us to improve the paper. 
We acknowledge the permission by Prof. Adam Riess to use the updated
MLCS template vectors for light curve fitting. Thanks are due to 
Konkoly Observatory and Calar Alto Observatory for 
generously allocating the necessary telescope time. 
The NASA Astrophysics Data
System, the SIMBAD database and the Canadian Astronomy Data
Centre were used to access data and references. 
The availability of these services are gratefully acknowledged. 

\end{acknowledgements}

\end{document}